\documentclass{aa}
\usepackage{graphicx}
\usepackage{amssymb}
\usepackage{color}
\begin{document}

\newcommand{\kms}{km s$^{-1}$}
\newcommand{\kmspc}{km s$^{-1}$ Mpc$^{-1}$}
\newcommand{\vhel}{$v_{hel}$}

\author{Fathi, K.\inst{1,2} \and Peletier, R. F.\inst{1,2}}
\institute{School of Physics \& Astronomy, University of
Nottingham, Nottingham, NG7 2RD, United Kingdom \and 
CRAL, Observatoire de Lyon, F-69561 St-Genis Laval cedex, France}

\offprints{\email{ppxkf@nottingham.ac.uk}}
\date{first submitted: Oct 2002, accepted: May 2003}
\title{Do Bulges of Early- and Late-type Spirals Have Different Morphology?}

\abstract{
We study HST/NICMOS H-band images of bulges of two equal-sized samples of 
early- ($T_{RC3} \leq 3$) and late-type spiral (mainly Sbc-Sc) galaxies 
matched in outer disk axis ratio. We find that bulges of late-type spirals 
are more elongated than their counterparts in early-type spirals. Using a 
KS-test we find that the two distributions are different at the 98.4\% 
confidence level. We conclude that the two data sets are 
different, i.e. late-type galaxies have a broader ellipticity 
distribution and contain more elongated features in the inner 
regions. We discuss the possibility that these would correspond to bars 
at a later evolutionary  stage, i.e. secularly evolved bars. Consequent 
implications are raised, and we discuss relevant questions regarding
the formation and structure of bulges. Are bulges of early-type and
late-type spirals different? Are their formation scenarios
different? Can we talk about bulges in the same way for different types 
of galaxies?

}

\maketitle

\section{Introduction}
\label{sec:intro}
Understanding bulges of galaxies is of pivotal importance for
improving our overall theories for galaxy formation and evolution.
Yet today, our general understanding of bulges is limited, and 
little is known about the formation scenario they have gone through. 
Up to the 1980's, ideas about bulge formation were very much influenced 
by Baade's population concept (e.g. Sandage 1986). 
Since in our Galaxy the stars in both the bulge and the stellar 
halo were found to be old, it was thought that the bulges and the
stellar galaxy halo were a single entity, with a common $r^{1 \over 4}$
surface brightness (SB) law (de Vaucouleurs 1948, 1959). They were thought to
be formed early on in the lifetime of the galaxy, during a rapid collapse 
phase (Eggen, Lynden-Bell \& Sandage 1962) before the formation of the disk. 
Their shapes were thought to be oblate spheroids (Kent, Dame \& Fazio 1991).
In recent years, however, a wealth of new data has altered our view of 
galactic bulges. Misalignments of the major axis with respect 
to the disk major axis indicate that bulges are probably not oblate 
(Bertola, Vietri \& Zeilinger 1988). Not all stars in the bulge region 
are old (e.g. B\"oker et al. 2002). And the surface brightness profiles 
of many bulges do not follow the $r^{1 \over 4}$ law - there are claims 
nowadays that there are no bulges for which the surface brightness 
profiles obey this law (Balcells et al. 2003). Bulges do seem to be 
rotationally flattened (Kormendy \& Illingworth 1982, Davies \& 
Illingworth 1983), with some bulges rotating even faster, almost 
as fast as disks (Kormendy 1993).

Before delving into the subject, it is good to first clearly define what we
mean by bulge. In this paper, by bulge we mean the photometric inner component
of the galaxy which appears to be superimposed onto an exponential disk. This
definition is extremely simple, but is at the same time very robust. The
definition is purely photometric, and is parametrical, but non-parametric
studies by e.g. Kent (1986) show that the bulge obtained in this way is
generally rounder than the disk, an indication that this definition is probably
measuring something physical. Note that the definition does not depend on
kinematics. When one measures the photometry of bulges defined in this way one 
finds that surface brightness profiles for galaxies of type earlier than Sc
rise more  steeply than those of late-types.  Surface brightness profiles
generally are    well described by  the S\'{e}rsic (1968) $r^{1 \over n}$
profile, for which the value of the  shape parameter $n$ varies from values
around 4 for bulges of  early-type spirals, to 1 for bulges of late-type
spirals (Andredakis, Peletier \& Balcells 1995; Phillips et al. 1996; Moriondo,
Giovanardi \& Hunt 1998). Bulges also host a variety of central components
including  dust lanes, star forming rings and spiral structure reaching  all
the way to the centre (Zaritsky, Rix \& Rieke 1993; Peletier et al.  1999;
Carollo et al. 2002).  Many bulges also contain central resolved sources,
identified as star clusters (Carollo et al. 2002, B\"oker et al. 2001), which
are fainter and less abundant  in early-type spiral galaxies.  Studying the
nuclear properties of bulges is important, since they generally scale with the
global galaxy properties (Faber et al. 1997). By studying the nuclear
properties of bulges of early and late-type spirals, one might be  able to
provide the relative importance of dynamical effects in forming and/or
maintaining the nuclear structure. It also provides a crucial test for
formation scenarios of spheroidal stellar systems along the entire luminosity
sequence. 

The main theories about bulge formation include scenarios such as: 
primordial collapse where bulge and disk form sequentially (Eggen, 
Lynden-Bell \& Sandage 1962); hierarchical galaxy formation (Kauffmann 
\& White 1993); galaxy mergers or infall of satellite galaxies (Zinn 
1985, Aguerri, Balcells \& Peletier 2001); and secular evolution 
(Kormendy 1979, Norman, Sellwood \& Hasan 1996, Hasan, Pfenniger \& 
Norman 1998), where central bars, formed by disk instabilities, thicken 
gradually to form bulges. The theories are still of crude, and 
theoretical predictions up to now have not been good enough for observers 
to rigorously test these scenarios.
Effects of flattening by rotation, or thickening of the
bars have not directly been verified in simulations for large numbers of 
objects, mainly due to lack of fast computers. There are a number of crude
predictions that one would naively draw from these formation models. 
If bulges were formed by secular evolution of bars, they would in general 
be flatter than if they were made in a primordial collapse. One might 
object against such simplified predictions, but since there is no 
alternative until robust results from simulations are available, we 
will assume that these predictions are correct.

In this paper, we investigate the ellipticity distribution of a 
large sample of early- and late-type spirals, to ultimately be able to infer
information about the formation scenario of bulges. Given that in the
infrared, the effects of dust extinction is considerably less 
than in optical, HST/NICMOS H-band data are
ideal for this task. We discuss a sub-sample of late-type spirals consisting 
of HST-archival data, together with a control sample of early-type spirals
consisting of HST-archival images of both Seyferts and non-Seyferts 
from Laine at al. (2002) (hereafter {\bf L02}). 
Since the NICMOS field of 
view is not large enough to provide Bulge-Disk decompositions, we use WFPC2
optical images for determination of the bulge radius, the radius where the 
light of the disk
starts dominating the galaxy light. Our samples are large
enough to allow us to draw statistical conclusions using the 
KS-test. The outline of the paper is as follows: In Sect. \ref{sec:data} we
introduce our data sets, and describe the reduction procedure. In Sect.
\ref{sec:analysis} the analysis procedure is outlined, and Sect.
\ref{sec:discussion} includes an extensive discussion of our results and
corresponding implications.

\section{The Data} 
\label{sec:data} 
We study a sample of archival HST/NICMOS
(NIC2) F160W ($\lambda_e \approx 1.6 \mu$ m, roughly 
corresponding to H-band) data.  The instrument and its 
performance are described in Thompson et al. (1998) and 
the NICMOS instrument handbook (MacKenty et al. 1997). 
For collecting the data the well known large observational 
data sets including samples by Mulchaey ID=7330, Peletier 
ID=7450, Pogge ID=7867, and Stiavelli ID=7331, were searched 
and investigated. These four samples are the only large 
samples of bulges in the HST archive, and include galaxies 
covering a wide range of morphological types. Of these 
samples, all galaxies of type later than or equal to 
$T_{RC3}= 4$ (corresponding to Sbc) are studied, 
and the earlier types are selected from the L02 sample.
The *\_cal.FITS images were used as a starting point, after 
which additional data reduction steps were performed. Processed 
by the CALNICA pipeline (NICMOS Data Handbook), these images 
have been science calibrated, dark subtracted, flat fielded, 
and cosmic ray corrected in an automatic way. Visual inspection 
of the images showed a number of anomalies, and it was evident 
that some further reduction needed to be implemented.

Sky subtraction was done by subtracting the mean value of 10 randomly chosen
outer regions of each image. Since in this study we are mainly  interested in
the central morphology, it is not important that the  galaxies are much larger
than the NIC2 frame. Masking bad pixels was done in two steps. First, only
the most evident regions were masked by fitting planes interpolated from the
boundary values (using our own software). This was followed by  fitting
ellipses to the isophotes (using {\small \sc Stsdas/Ellipse}),  and
investigating the residual images, which showed the presence of additional
undesired regions, such as very bright star forming knots,  foreground stars,
or very strong dust regions. After further masking of these regions, all the
obvious artifacts were removed, and the images were ready for final ellipse
fitting.  The {\small \sc Ellipse} task in the {\small \sc Isophote} package
in  {\small \sc Stsdas} was used on the masked images, this time providing more
accurate results. The isophotes of each galaxy were fitted with a sequence of
ellipses with different semi-major axis lengths based on the algorithm 
described by Jedrzejewski (1987). The final ellipses were then deprojected,
using a two-dimensional deprojection procedure assuming that the galaxy was a 
thin disk, with inclination (given by the axis ratio in the outer parts) and PA 
inferred from the RC3 catalogue, (similar to the method used in L02).
The deprojected fits are hereafter used in our analysis. Although removing the 
bad areas, before fitting ellipses, is believed to not seriously 
affect the fitting procedure (Carollo, Stiavelli  \& Mack 1998), 
we can be sure that the statistics are not affected by masking bad regions, 
since both our samples are treated in the same manner.

\subsection{Sample Selection} 
\label{sec:sample} 
We decided to split the sample of bulges at $T_{RC3} = 4$, since our
previous studies have shown that bulges of spirals with types earlier
than that have several properties in common with ellipticals, such as
surface brightness (SB) profile, age, age spread, and agreement with the
fundamental plane of ellipticals (eg. de Jong 1996; Peletier et al. 1999; 
Falc\'on-Barroso, Peletier \& Balcells 2002).
Bulges of late-type spirals have been shown to be different. We require
that the galaxies are not too inclined, as this would imply large projection
effects, and hide important features. Absolute $B$ magnitude, and distance (or
\vhel ) requirements are imposed, in order to avoid large differences in
spatial resolution, and/or discrepant prominence of small scale
structures such as stellar clusters or dust lanes. Small scale features are
more prominent for nearby galaxies, whereas for more distant galaxies, we 
would not be able to see much of the inner parts. 
The same criteria were used when selecting the early-type sample from L02, 
for which in addition we also rejected all the Seyfert 1 galaxies to avoid
contamination of the central regions by the bright nucleus.
\begin{figure*}[hbt]
\label{fig:sample_comparison} 
\includegraphics[width=18cm, angle=0]{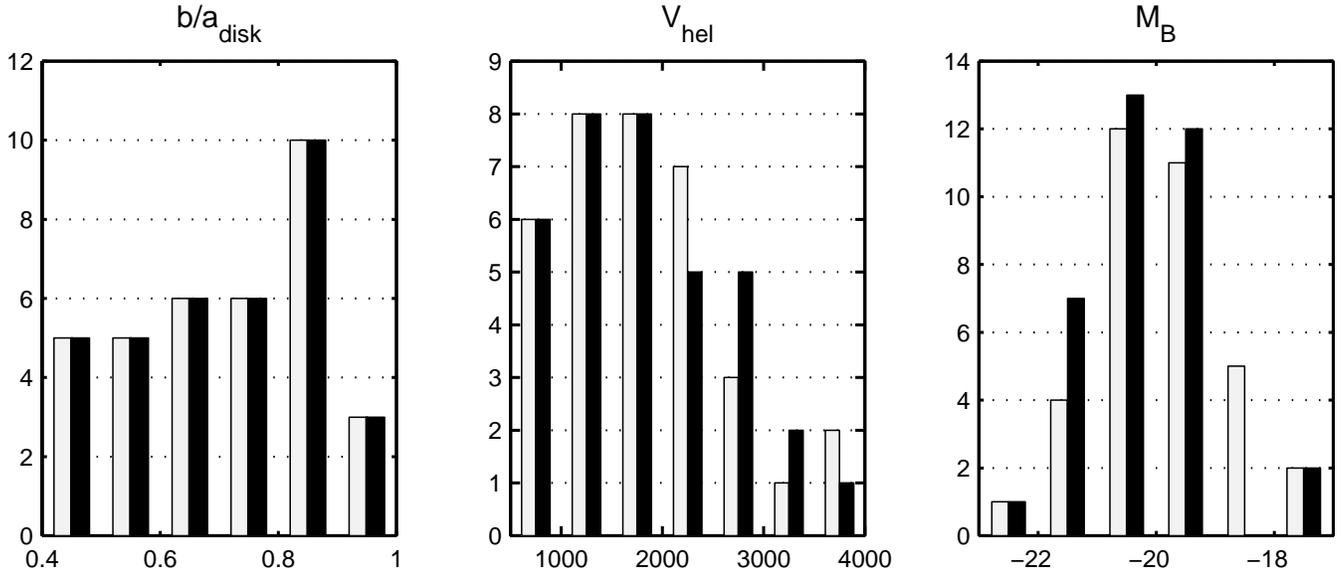}
\caption{Distribution histograms of $b/a$, \vhel, and $M_B$ of the 
two samples, The black colour represents the early-type control 
sample and the lighter colour represents the late-type sample. Each 
subsample contains 35 galaxies, all of which fulfil all the  selection
criteria described in the text above, and for obtaining the $M_B$,
$H_0 = 70$ \kmspc was used.} 
\end{figure*}
\begin{table}
\caption{List of 35 rejected late-type galaxies and reasons
for rejection (see text for further details). The remaining 35 galaxies
fulfill all our criteria and match well with the early-type sample of L02.}
\label{tab:rejectedlist}
\begin{center}
{\tabcolsep=10.pt
\begin{tabular}{lcl}\hline
Galaxy & $T_{RC3}$ & Reason for rejection               \\ \hline \hline
ESO 290-26 &   4 &   Ellipse fitting fails  (faint)	\\
ESO 443-80 &   9 &   Bulge never dominates 		\\
ESO 499-37 &   7 &   Ellipse fitting fails (faint) 	\\
ESO 549-18 &   5 &   Ellipse fitting fails (faint) 	\\
ESO 549-2  &   9 &   Ellipse fitting fails (faint) 	\\
ESO 572-22 &   7 &   Bulge never dominates 		\\
IC 1555	   &   7 &   Ellipse fitting fails (patchy) 	\\
NGC 151	   &   4 &   No B/D decomposition 		\\
NGC 406	   &   5 &   Bulge never dominates 		\\
NGC 578	   &   5 &   No B/D decomposition 		\\
NGC 1483   &   4 &   Bulge never dominates 		\\
NGC 1800   &   9 &   Ellipse fitting fails (patchy)	\\
NGC 1892   &   6 &   Bulge never dominates 		\\
NGC 2104   &   9 &   Bulge never dominates 		\\
NGC 2336   &   4 &   Matching with L02 sample 		\\
NGC 2748   &   4 &   Bulge never dominates 		\\
NGC 2964   &   4 &   Matching with L02 sample 		\\
NGC 3079   &   5 &   Bulge never dominates 		\\
NGC 3259   &   4 &   Matching with L02 sample 		\\
NGC 4536   &   4 &   Matching with L02 sample 		\\
NGC 5054   &   4 &   Matching with L02 sample 		\\
\hline
\end{tabular}}
\end{center}
\end{table}

An initial sample of all bulges included 70 objects (all late-type), 
on which we applied further selection. Using
parameters from the RC3 catalogue (de Vaucouleurs et al. 1991), we removed all
the galaxies more inclined than $\simeq 63^\circ$ corresponding to the apparent
outer, $R_{25}$ disk axis ratio of $b/a \leq 0.45$. Subsequently, the sample
was restricted to a limited radial velocity range of 500 (km s$^{-1}$) $\leq$ \vhel 
$\leq$ 4000 (\kms), and absolute B-magnitude of $M_B \leq -15$
(all inferred from the RC3 catalogue). 
As we needed to estimate the bulge radius by performing a bulge-disk 
decomposition (see Sect. \ref{sec:decomposition}) on the corresponding 
WFPC2 images, we required availability of WFPC2 optical images on which 
ellipse fitting was possible. 
This rules out off-field galaxies (i.e. galaxies for which the central 
areas were only partly on the field), 
low SB galaxies and those with very strong star forming
regions and/or strong dust features. Although the H-band images 
are believed to be unaffected by these effects, 
a few galaxies had to be removed for this purpose. 
Table \ref{tab:rejectedlist} quantifies the number of rejected 
galaxies after application of each selection
criterion. Finally, to obtain reliable statistics, 
it is of vital importance that the two samples are 
similarly distributed in all salient parameters.
We compare our sample with the sample of L02 to make a matching control
sample of early type spiral bulges, applying the same ranges in $M_B$,
distance, and $i$, and since not many galaxies with
$0.4<b/a<0.6$ are available in the sample of L02 we removed 5
galaxies from our late-type sample. Given that the very bright nucleus 
of type 1 Seyfert galaxies affects the 
central parts of the images, the Seyferts selected from 
the L02 sample are all of type 2.
The final remaining list of the two subsamples are found in 
Tables \ref{tab:latesample} \& \ref{tab:earlysample}, and the 
corresponding distributions are illustrated in Fig. 1.
\begin{table*}[hbt]
\caption{List of 35 remaining late-type galaxies. 
The first four columns indicate galaxy name, morphological Type, 
outer disk axis ratio $b/a$,  and radial velocities from the RC3 
catalogue. The three ellipticity columns  are the derived projected, 
deprojected, and average bulge ellipticities as described in the text, 
and the derived $B/T$ ratio using equation (\ref{eq:bt}). The last 
column is the bulge radius obtained from bulge-disk decomposition of 
the WFPC2 images. The $\dag$ marks galaxies for which a nuclear bar 
has been detected, i.e. we have found an ellipticity variation greater 
than 0.1 within the bulge region.}
\label{tab:latesample}
\begin{center}
{\tabcolsep=11.pt
\begin{tabular}{lllrccccc}\hline 
\multicolumn{1}{c}{Galaxy}  & Type &  $b/a$  & $cz$ (\kms)  &  $\epsilon_{cenPR}$ 
& $\epsilon_{cenDEP}$ & $\epsilon_{average}$ & $B/T$ & $r_B$ (arcsec)\\ \hline  \hline
ESO 404-G3 $^\dag$&  .SBT4P.&   0.50 & 2383  &  0.65   &   0.40 &   0.35 &   0.013&  0.80  \\
ESO 498-G5 	  &  .SXS4P.&   0.81 & 2413  &  0.10   &   0.15 &   0.15 &   0.024&  1.86  \\
IC 5273    	  &  .SBT6*.&   0.66 & 1206  &  0.72   &   0.65 &   0.50 &   0.011&  1.74  \\
NGC 289    	  &  .SBT4..&   0.71 & 1690  &  0.30   &   0.10 &   0.15 &   0.009&  1.14  \\
NGC 1300   	  &  .SBT4..&   0.66 & 1592  &  0.10   &   0.30 &   0.30 &   0.030&  2.29  \\
NGC 1345 $^\dag$  &  .SBS5P*&   0.74 & 1543  &  0.70   &   0.40 &   0.45 &   0.394&  9.15  \\
NGC 1688 $^\dag$  &  .SBT7..&   0.78 & 1223  &  0.60   &   0.60 &   0.50 &   0.323&  0.88  \\
NGC 1961   	  &  .SXT5..&   0.65 & 3983  &  0.25   &   0.40 &   0.35 &   0.009&  1.86  \\
NGC 2276   	  &  .SXT5..&   0.96 & 2372  &  0.25   &   0.25 &   0.20 &   0.005&  0.78  \\
NGC 2339 $^\dag$  &  .SXT4..&   0.76 & 2361  &  0.35   &   0.45 &   0.30 &   0.025&  2.15  \\
NGC 2344   	  &  .SAT5*.&   0.98 &  914  &  0.05   &   0.05 &   0.05 &   0.045&  1.89  \\
NGC 2903   	  &  .SXT4..&   0.48 &  565  &  0.22   &   0.40 &   0.40 &   0.003&  1.18  \\
NGC 3145 $^\dag$  &  .SBT4..&   0.51 &  3656 &  0.35   &   0.55 &   0.40 &   0.062&  2.54  \\
NGC 3949          &  .SAS4*.&   0.58 &   681 &  0.35   &   0.25 &   0.25 &   0.012&  0.75  \\
NGC 4030 $^\dag$  &  .SAS4..&   0.72 &  1449 &  0.15   &   0.30 &   0.25 &   0.025&  2.59  \\
NGC 4303 $^\dag$  &  .SXT4..&   0.89 &  1607 &  0.30   &   0.35 &   0.15 &   0.002&  2.44  \\
NGC 4806 $^\dag$  &  .SBS5?.&   0.83 &  2430 &  0.20   &   0.25 &   0.25 &   0.004&  0.91  \\
NGC 4939   	  &  .SAS4..&   0.51 &  3091 &  0.35   &   0.25 &   0.25 &   0.037&  2.72  \\
NGC 5005   	  &  .SXT4..&   0.48 &   992 &  0.55   &   0.50 &   0.20 &   0.078&  1.95  \\
NGC 5033 $^\dag$  &  .SAS5..&   0.47 &   861 &  0.45   &   0.40 &   0.30 &   0.035&  1.63  \\
NGC 5427   	  &  .SAS5P.&   0.85 &  2645 &  0.05   &   0.15 &   0.10 &   0.017&  1.61  \\
NGC 5643   	  &  .SXT5..&   0.87 &  1163 &  0.10   &   0.15 &   0.10 &   0.052&  1.07  \\
NGC 6000 $^\dag$  &  .SBS4*.&   0.87 &  2110 &  0.55   &   0.55 &   0.25 &   0.055&  1.65  \\
NGC 6217 $^\dag$  &  RSBT4..&   0.83 &  1368 &  0.35   &   0.45 &   0.25 &   0.141&  2.56  \\
NGC 6221 $^\dag$  &  .SBS5..&   0.69 &  1350 &  0.35   &   0.40 &   0.20 &   0.027&  2.40  \\
NGC 6384   	  &  .SXR4..&   0.66 &  1690 &  0.30   &   0.15 &   0.15 &   0.079&  2.37  \\
NGC 6412 $^\dag$  &  .SAS5..&   0.87 &  1475 &  0.34   &   0.45 &   0.20 &   0.006&  1.87  \\
NGC 6744   	  &  .SXR4..&   0.65 &   730 &  0.20   &   0.25 &   0.25 &   0.016&  1.70  \\
NGC 6814          &  .SXT4..&   0.93 &  1509 &  0.05   &   0.10 &   0.10 &   0.054&  1.98  \\
NGC 6951   	  &  .SXT4..&   0.83 &  1331 &  0.19   &   0.15 &   0.15 &   0.030&  2.13  \\
NGC 7126   	  &  .SAT5..&   0.46 &  2980 &  0.30   &   0.40 &   0.40 &   0.057&  2.24  \\
NGC 7188   	  &  PSBT4..&   0.47 &  1767 &  0.15   &   0.45 &   0.40 &   0.009&  0.57  \\
NGC 7392   	  &  .SAS4..&   0.59 &  2908 &  0.34   &   0.25 &   0.25 &   0.018&  2.13  \\
NGC 7421   	  &  .SBT4..&   0.89 &  1830 &  0.20   &   0.20 &   0.20 &   0.027&  1.44  \\
NGC 7479   	  &  .SBS5..&   0.76 &  2394 &  0.35   &   0.20 &   0.20 &   0.023&  1.12  \\
\hline
\end{tabular}}
\end{center}
\end{table*}
\begin{table*}
\caption{List of 35 remaining early-type galaxies. The
columns are the same as in Table \ref{tab:latesample}.
Error values for the ellipticities are of order $<0.002$. For the B/T 
ratio, the errors are of order $<0.005$, and for $r_B$ the errors are 
of order $<0.5$ arcsec. (Note that the errors 
for Table \ref{tab:latesample} are of same order.)}
\label{tab:earlysample}
\begin{center}
{\tabcolsep=9.pt
\begin{tabular}{lllrccccc}\hline 
\multicolumn{1}{c}{Galaxy}   & Type &   $b/a$  &  $cz$ (\kms)   & $\epsilon_{cenPR}$ 
& $\epsilon_{cenDEP}$ 	& $\epsilon_{average}$ & $B/T$  & $r_B$ (arcsec) \\ \hline \hline
ESO 137 G-34$^\dag$     &.SXS0?.  & 0.76   &  2620 &   0.15  & 0.15 & 0.15 & 0.013 & 3.80 \\
IC 2560 $^\dag$ 	&PSBR3*.  &  0.63  &  2873 &   0.25  & 0.35 & 0.15 & 0.020 & 1.77 \\
NGC 1365        	&.SBS3..  &  0.55  &  1675 &   0.20  & 0.55 & 0.40 & 0.008 & 1.10 \\
NGC 1530 $^\dag$ &.SBT3.. &  0.52  &  2506 &  0.20  & 0.60 &  0.40 &  0.019  & 3.13\\
NGC 1672 $^\dag$	&.SBS3..  &  0.83  &  1282 &   0.15  & 0.20 & 0.10 & 0.110 & 1.76 \\
NGC 2460 	 &.SAS1.. &  0.76  &  1442 &  0.25  & 0.20 &  0.20 &  0.027  & 2.13\\
NGC 2639 		&RSAR1*\$ &  0.60  &  3198 &   0.20  & 0.20 & 0.20 & 0.032 & 2.22 \\
NGC 3032 	 &.LXR0.. &  0.89  &  1568 &  0.10  & 0.05 &  0.05 &  0.037  & 1.76\\
NGC 3081 		&RSXR0..  &  0.78  &  2391 &   0.20  & 0.30 & 0.30 & 0.027 & 2.38 \\
NGC 3169 $^\dag$ &.SAS1P. &  0.63  &  1261 &  0.25  & 0.25 &  0.20 &  0.020  & 1.95\\
NGC 3227 		&.SXS1P.  &  0.68  &  1145 &   0.20  & 0.20 & 0.20 & 0.028 & 1.65 \\
NGC 3277 	 &.SAR2.. &  0.89  &  1460 &  0.10  & 0.05 &  0.05 &  0.470  & 0.56\\
NGC 3300 	 &.LXR0*\$&  0.52  &  3045 &  0.30  & 0.35 &  0.35 &  0.025  & 2.35\\
NGC 3982 		&.SXR3*.  &  0.87  &   924 &   0.20  & 0.10 & 0.10 & 0.015 & 3.58 \\
NGC 4117 		&.L..0*.  &  0.49  &   871 &   0.30  & 0.25 & 0.25 & 0.008 & 2.21 \\
NGC 4143 	 &.LXS0.. &  0.63  &   784 &  0.20  & 0.20 &  0.20 &  0.105  & 2.13\\
NGC 4151 $^\dag$	&PSXT2*.  &  0.71  &   956 &   0.20  & 0.20 & 0.20 & 0.068 & 1.53 \\
NGC 4260 	 &.SBS1.. &  0.50  &  1886 &  0.30  & 0.30 &  0.30 &  0.008  & 2.36\\
NGC 4384 	 &.S..1.. &  0.78  &  2400 &  0.60  & 0.60 &  0.60 &  0.084  & 4.18\\
NGC 4725 		&.SXR2P.  &  0.71  &  1180 &   0.10  & 0.25 & 0.25 & 0.047 & 2.12 \\
NGC 4941 $^\dag$	&RSXR2*.  &  0.54  &   846 &   0.40  & 0.30 & 0.25 & 0.009 & 2.21 \\
NGC 5064         &PSA.2*. &  0.46  &  2952 &  0.50  & 0.20 &  0.20 &  0.046  & 2.36\\
NGC 5273 		&.LAS0..  &  0.91  &  1054 &   0.15  & 0.05 & 0.05 & 0.020 & 3.04 \\
NGC 5377 	 &RSBS1.. &  0.56  &  1830 &  0.30  & 0.30 &  0.30 &  0.026  & 2.13\\
NGC 5383 $^\dag$ &PSBT3*P &  0.85  &  2226 &  0.25  & 0.20 &  0.10 &  0.014  & 2.58\\
NGC 5448 $^\dag$ &RSXR1.. &  0.46  &  1973 &  0.40  & 0.20 &  0.20 &  0.023  & 2.13\\
NGC 5614 	 &.SAR2P. &  0.83  &  3872 &  0.05  & 0.20 &  0.20 &  0.078  & 1.60\\
NGC 5678 $^\dag$ &.SXT3.. &  0.49  &  2267 &  0.30  & 0.25 &  0.25 &  0.015  & 1.77\\
NGC 5953 		&.SA.1*P  &  0.83  &  2099 &   0.20  & 0.25 & 0.25 & 0.127 & 2.17 \\
NGC 6300 $^\dag$	&.SBT3..  &  0.66  &  1064 &   0.30  & 0.30 & 0.20 & 0.006 & 2.93 \\
NGC 7496        	&.SBS3..  &  0.91  &  1527 &   0.40  & 0.35 & 0.20 & 0.016 & 2.45 \\
NGC 7217 	 &RSAR2.. &  0.83  &   935 &  0.05  & 0.10 &  0.10 &  0.026  & 2.58\\
NGC 7716 	 &.SXR3*. &  0.83  &  2541 &  0.30  & 0.25 &  0.15 &  0.028  & 3.79\\
NGC 7742 $^\dag$ &.SAR3.. &  1.00  &  1661 &  0.10  & 0.10 &  0.05 &  0.031  & 1.61\\
NGC 7743 		&RLBS+..  &  0.85  &  1722 &   0.10  & 0.15 & 0.15 & 0.019 & 2.61 \\
\hline	
\end{tabular}}		
\end{center}
\end{table*}

\subsection{Bulge-Disk Decomposition} 
\label{sec:decomposition} 
A very important quantity for our study is the ``bulge radius''. 
This is defined as the radius at which the SB of the exponential 
disk equals the surface brightness of the bulge. Successful fits 
to the bulges of spiral galaxies have been obtained using the Hubble 
law (Hubble 1930), King model (King 1966), de Vaucouleurs $r^{1\over 4}$ 
law  (de Vaucouleurs 1948), by a generalised version of de Vaucouleurs' 
law $r^{1 \over n}$ (Caon, Capaccioli \& D'Onofrio 1993; Andredakis et 
al. 1995) and by an exponential function (Kent et al. 1991; Andredakis 
\& Sanders 1994; Baggett, Baggett \& Anderson 1998). Andredakis \& 
Sanders (1994) and Andredakis et al. (1995) showed that for the late-type 
spirals of this paper an exponential fit of the bulge is entirely adequate, 
and much better than using an r$^{1 \over 4}$ law. 

We decomposed the galaxy SB profiles, fitting exponentials to 
both the bulge and the disk part. Despite the fact that 
potentially better fits can be obtained with a S\'ersic $r^{1 \over n}$
law for the bulge, we decided to use exponential distributions, since these 
produce the most robust determinations for the bulge radius. As a test we
have compared bulge radii obtained using a S\'ersic + exponential 
distribution with bulge radii obtained with double-exponential 
distributions and found that both values are not very different 
in log-scale (see Table 4). We find that the measured bulge 
ellipticities do not change significantly when S\'ersic bulge 
radii are used and our statistical conclusions are not affected. 
Since the field of view of 
\begin{table}[hbt]
\caption{Average bulge radii (in arc seconds) with corresponding RMS 
(in brackets) obtained from fitting S\'ersic + exponential and 
double-exponential SB profiles to the WFPC2 images.}
\label{tab:rejectedlist}
\begin{center}
{\tabcolsep=6.pt
\begin{tabular}{lcc}\hline
 & $\overline{r_B}$ [S\'er + Exp] & $\overline{r_B}$ [Exp + Exp]\\ \hline \hline
Early-types &   6.04 (3.27) &   2.30 (0.76)	\\
Late-types  &   3.80 (2.16) &   1.95 (1.40) 	\\
\hline
\end{tabular}}
\end{center}
\end{table}

NICMOS is too small to be able to perform the Bulge-Disk 
decomposition on, we used optical WFPC2 images for this purpose. 
The optically obtained bulge radii do not differ much from those 
from the H-band images (Carollo et al. 2002). Images with the 
F555W, F547M, F606W, and F814W filters were used with the galaxy 
nucleus generally on the Planetary Camera chip. A manual reduction 
(bad pixel masking, sky subtraction etc.) as described previously 
was performed on the mosaiced images, followed by ellipse fitting. 
As a result of the high sensitivity of the optical images to dust extinction, 
the axis ratio of the successive ellipses was fixed 
(axis ratio and PA as used for the deprojection) throughout the 
galaxy. The resulting average radial SB profile is then decomposed 
into bulge and disk, and the bulge radius is obtained.

\section{Analysis} 
\label{sec:analysis} 
\begin{figure*}[hbt]
\label{fig:deprojected}
\includegraphics[width=18cm, angle=0]{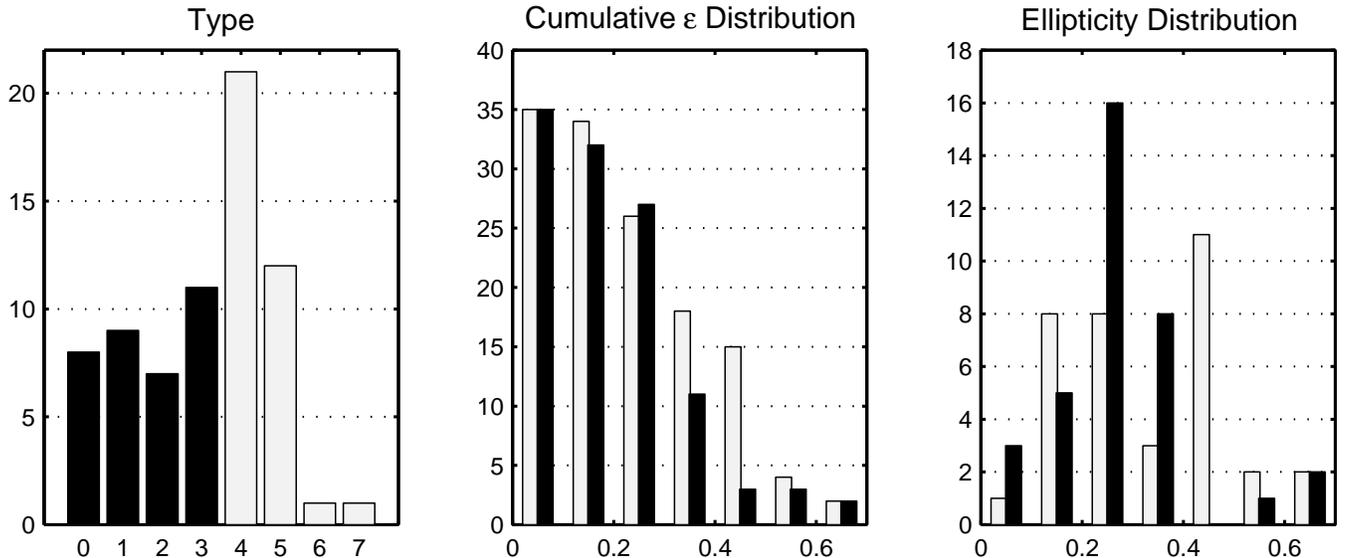}
\caption{Histograms of type distribution, deprojected bulge 
ellipticity $(\epsilon \equiv 1-{b \over a})$ cumulative distribution, 
and bulge ellipticity distribution  (in order from left to right) of 
our late-type and early-type samples. The black colour represents the 
early-type subsample.} 
\end{figure*}
The reduced data are now ready to be analysed. In Fig. 7 we plot the 
deprojected ellipticities and PAs,  obtained from the ellipse fitting, 
as a function of radius (Fig. 7). We used the bulge radii, from the 
bulge-disk decomposition, to estimate the extent of the bulge region. In 
this region we quantified the bulge ellipticities in the following way: 
When the ellipticity profile within the bulge varies by less than 0.1, 
we define the bulge  ellipticity as the mean value. Alternatively, 
when there is a rise or fall in the  ellipticity of more than 0.1, 
we take the peak ellipticity. These two definitions provide the bulge 
ellipticity for the whole of our sample (see the histograms in bins of 
0.1 in Fig. 2). As we are comparing the results with that of a control 
sample, any conclusion investigating whether the two datasets are 
statistically significant needs to be based on a statistical test 
of the two samples. 

We used the Kolmogorov-Smirnov KS-test, which has the advantage 
of making no initial assumption about the distribution of 
data, i.e. it is non-parametric and distribution free 
(Press et al. 1992). For our deprojected ellipticity distributions 
we obtain a KS probability value of $0.0160$, meaning that the two 
distributions are different at the 98.4\% confidence level. 

To find out whether this result is real we applied a number of tests. 
We started by comparing the projected ellipticities of early and late 
type spirals. The corresponding distributions of the projected bulge 
ellipticities are given in Fig. 3. The two distributions are peaked 
at different values (e.g. there is a significant excess of late-type 
galaxies with bulge ellipticities in the $0.3-0.4$ bin, and 
a deficit for the $0.2-0.3$ bin).  
The bulges of late-type spirals are more flattened even when 
comparing the projected ellipticities, although in 
this case the KS-test leads to a probability
value of $0.0630$, corresponding to 93.7\% confidence level. 
The projected ellipticities are consistent with a single distribution,
not necessarily implying the same after deprojection.
\begin{figure}[hbt]
\label{fig:projected} 
\resizebox{\hsize}{!}{
\includegraphics[width=4cm]{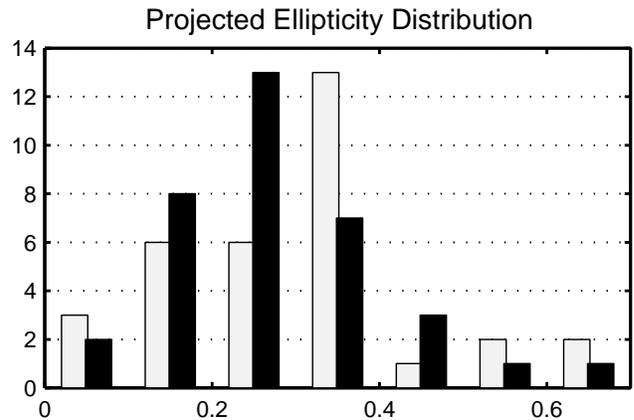}}
\caption{The distribution of projected bulge ellipticities 
for our early- and late-type samples. The black colour 
represents the early-type sample.}
\end{figure}

We continue by defining the bulge ellipticity in a different
way, namely such that in all the cases we would take this value to 
be the average ellipticity of the bulge component of the galaxy. 
This provides a different distribution (see Fig. 4), 
with a KS-test result of $0.3200$. This implies that the 
ellipticity distribution of bulges in early and late type spirals is 
similar, and that the only difference between the two groups is the
structure inside the bulge.

Investigating the residual images after fitting ellipses (Fig. 8) shows 
that in 12 cases the residual images are very smooth all the way to the 
very centre of the galaxy. This means that the galaxy light is very well 
described by the set of fitted ellipses This fact shows that the galaxy 
does not contain very much dust, and that the ellipticity analysis is a 
very reliable indicator of the distribution of the light. 
In the remaining 23 cases, the 
residual images show strong central features meaning that there are
significant structures in the central parts. These are disks, bars,
strong star formation regions, spiral structure, and/or dust features. 
In these cases the residual images also show clear indications of dusty 
bulges for which the inferred morphology is affected by the dust 
(e.g. the position of the centre changes as a function of radius). 
\begin{figure}
\label{fig:average}
\resizebox{\hsize}{!}{
\includegraphics[width=4cm]{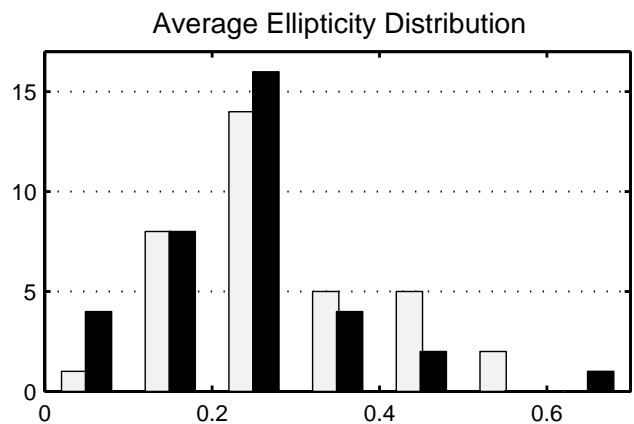}}
\caption{Bulge ellipticity distribution for the average $\epsilon$
values. Colour convention is same as before.}
\end{figure}

\section{Discussion} 
\label{sec:discussion} 
\subsection{Possible Observational Bias}
\label{sec:bias}
We have found that the distribution of bulge ellipticities is different
between early and late-type spirals (see Fig. 2). The KS-test indicates 
however that the statistical significance of this result is small. 
Could it be that we made unjustified assumptions during the deprojection?
We discuss here a number of observational biases which could cause this 
result and find out whether they might be significant.

The difference between the ellipticity distributions might be caused by the 
inferred 2D deprojection. This procedure assumes that the galaxy is two 
dimensional, which in reality is not the case. Depending on the thickness
variations along the galaxy plane the ``deprojected'' bulge may seem more 
elongated as an artifact of the 2D deprojection. If the bulge is much thicker 
than the underlying disk it becomes more elongated, and consequently if bulges 
of late-type galaxies are intrinsically thicker than those of early-types, our 
statistics could be biased. For this purpose we analysed the intrinsic bulge 
thickness of a sample of 20 edge-on galaxies in the K-band and in the same 
Hubble-type range as our main sample (Guijarro Roman et al., in preparation). 
The sample consists of 14 early-type spiral galaxies and 6 late-type spirals, 
a subset of the sample of de Grijs (1998). For the 20 galaxies we derived the 
bulge scale heights by directly fitting exponentials to the vertical light 
profile, and the bulge scale length through B/D decomposition of the radial 
profile (as described in Sect. \ref{sec:decomposition}). The ratio of the bulge 
scale height ($h_z$) to the bulge scale length ($h_r$) directly measures the 
bulge axis ratio, and a comparison of these value tells whether the bulge of 
one galaxy is flatter or thicker than another. Fig. 5 shows that this ratio 
is not dependent on type, and that bulges of early-type spirals are not 
intrinsically thicker than bulges of late-type spirals, or vice versa. 
Hence, the 2D deprojection does not add any bias to our statistics for 
comparing the deprojected ellipticities.
\begin{figure}
\label{fig:scaleratio}
\resizebox{\hsize}{!}{
\includegraphics[width=4cm]{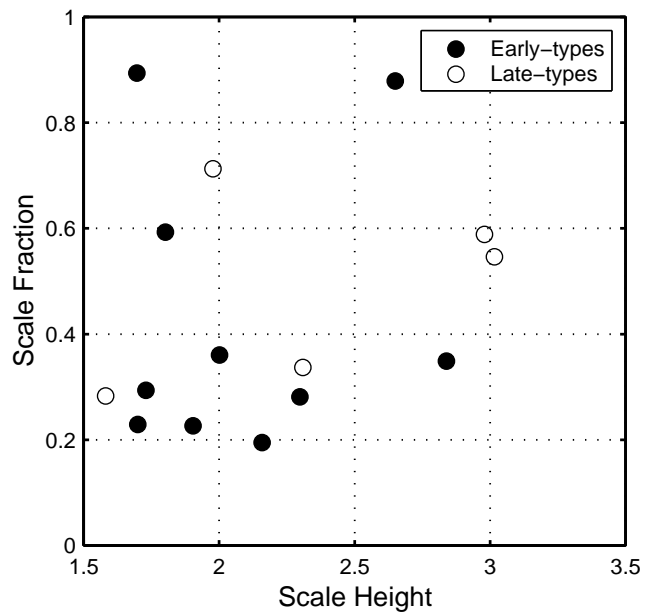}}
\caption{$h_z \over h_r$ vs. $h_z$ for 15 of the sample edge on galaxies. 
Filled black circles are the values for the early-type spirals. It is evident 
that intrinsic thickness is independent of galaxy type. Note also that 5 of 
our initial sample edge ons did not show to have a prominent bulge, and therefore 
were not taken into consideration, as the obtained $h_z \over h_r$ vs. $h_z$ 
would be unreliable.}
\end{figure}

It could be that the difference is caused by the fact that 
it is more difficult to detect elongated structures in big bulges 
of early-type spirals. We can investigate that by looking at the 
frequency of bars detected inside the bulge regions. 
Defining bars in the same way as in Knapen et al. (2000) we find no 
difference in the frequency of bars in the bulge region between 
early- and late-type galaxies. In our sample, 12 early-type and 
13 late-type spirals indicate the presence of a bar within the bulge. 
These numbers are very similar, and no discrepancy is observed. 

Instead of galaxy type we might also look at the 
bulge to total (B/T) flux ratio, given by
\begin{equation}
\label{eq:bt}
\frac{B}{T} = 
\frac{I_{0B} \times h_{B}^{2}}{I_{0B} \times h_{B}^{2} \; + \; I_{0D} \times h_{D}^{2}},
\end{equation}
where $I_{0B}$ and $h_{B}$ denote the central intensity and the scale
length of the bulge, respectively, and the index $D$ indicates the same
parameters for the disk component.

We derive this parameter for our entire sample and divide the sample 
in two equal-sized subsamples according to this value, drawing the 
middle line. This results in two subsamples, one containing galaxies 
with B/T ratio $<0.026$, and one containing the rest of the sample. 
The bulge peak ellipticities are then compared (see Fig. 6), and the
corresponding KS-test probability value is $0.1975$. As also seen from the
histogram, one cannot say that the distributions are different.
Although one might think that the B/T ratio is a good indicator of galaxy 
type, it is not a very good parameter for our sample here, since most B/T 
ratios are very small (less than 0.05), and in this range of B/T it is not 
a very good morphological type indicator.
\begin{figure}
\label{fig:BTratio}
\resizebox{\hsize}{!}{
\includegraphics[width=4cm]{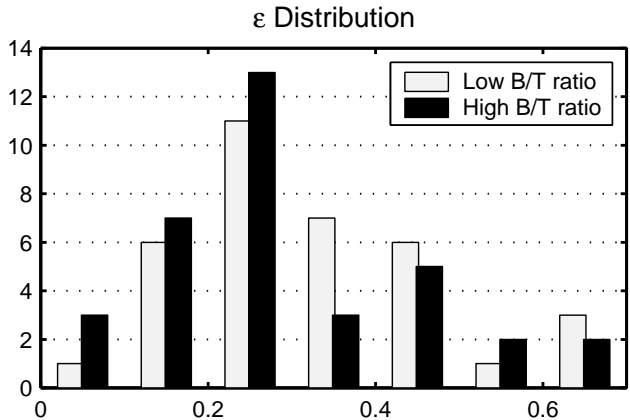}}
\caption{Distribution of deprojected ellipticities as a function of B/T ratio. The median B/T ratio is $B/T=0.026$.}
\end{figure}

\subsection{Implications}
\label{Implications}
We find that bulges of late-type galaxies are more elongated. What does this
mean for the formation of bulges? Are we seeing some signs of secular
evolution here? Could it be that secular evolution destroys nuclear bars, 
converting them slowly into bulges, which as a result are still somewhat more 
elongated?

Theoretical studies have shown that bars are capable of evolving 
in self dissolving mechanisms (Sellwood \& Wilkinson 1993), i.e  
they undergo secular evolution (Toomre 1966, Hasan \& Norman 1990;
Pfenniger \& Norman 1990). Secular evolution, 
however, is not always believed to destroy the bars. Merritt \&
Sellwood (1994) showed that the thickening of bars, through off plane bending
instabilities, stops when the density in the mid-plane drops to a low enough
value that the natural vertical frequency for a large fraction of particles
drops below the forcing frequency below the global bend. There is also 
some strong theoretical evidence that long lived and strong bars 
survive loss of up to 2/3 of the angular momentum and reduced 
pattern speed by a factor of up to 5 (Sellwood \& Debattista 1996).

Spiral activity has been shown to occur for a longer time in 
the outer disk when the initial bar ends at a radius well 
inside the outer edge of the disk. This spiral activity in the 
disk outside the bar is not affected by the bar, although the 
bar grows by trapping additional particles which are ready to 
lose angular momentum near the inner ends. The new 
particles added to the bar in this way still have 
too much angular momentum to sink deep into the bar, 
and are therefore added to the outer ends of the bar. 
As a result, the angular momentum of the bar content 
is rising although the bar itself is slowing down. 
Bar destruction and pattern speed change are also explained 
with other theories. Interaction with the halo (Weinberg 1985, 
Debattista \& Sellwood 1996, Athanassoula 1996) is believed to 
be able to slow down a bar through dynamical friction. Mergers 
(Gerin, Combes \& Athanassoula 1990; Athanassoula 2002; Heller 
\& Shlosman 1994; Barnes \& Hernquist 1992), or fuelling of an 
active nucleus by driving gas towards the centre, play a considerable 
role in destruction of bars resulting in a spheroidal bulge 
(Norman, Sellwood \& Hasan 1996). 

It is not obvious that secular evolution of bars leads to oblate structures.
Secularly evolved bars may also end up as elongated structures, as 
bars are believed to be populated by orbits of greater eccentricity 
than those in the axisymmetric parts of the disk.
One might think that bar formation, on the other hand, is expected to be less
common in  early-type systems (Kormendy 1979, 1993; Combes \& Elmegreen 1993, 
Sellwood \& Wilkinson 1993; L\"utticke, Dettmar \& Pohlen 2000), as  there is
less gas in early-type spirals, and there is also already  a big bulge which
stabilises the galaxy and prevents  self-gravitating bar formation. If secular
evolution is active one would therefore expect to find more elongated 
structures in late-type spirals, as we find observationally.
This argument, however, might not be correct, since recent statistics on 
bars (e.g. L\"utticke et al. 2000) shows that the bar fraction in early 
type spirals (Sa's) is the same as in Sc's. Also, Athanassoula (2002) has 
recently shown that massive bulges do not always suppress bar formation, 
when more accurate simulations with live-particles bulges/halos are done. 
In any case, in this paper we do not see a big difference between shapes 
of central features in early and late type spirals, so the process which 
makes elongated bulges in late-type spirals is not much more efficient 
than in early-type spirals.

To summarise, secular evolution seems to be able to make bulges 
(central disk concentrations) from bars, but it is not clear what 
their final shape is, and how efficient this process is.
Yet so far it is not possible to confirm nor refute any of these 
theories, as very few observational studies have been done (Gerssen 
et al. 1999, Zimmer \& Rand 2002), and only improved observations will 
be able to tell whether secular evolution of bulges is indeed possible.

\section{Summary}
\label{sec:summary}
Studying a sample of 70 galaxies in the H-band, we find that 
inner features in bulges of late-type spirals are more elongated than 
those in early-type spirals, although the statistical significance of 
this result is small ($P_{KS} = 0.0160$). 
When we simply compare the average ellipticity in the bulge between both 
samples, we don't find any difference. This probably indicates that bulges 
of later type spiral galaxies contain more elongated features like nucluear 
bars than bulges of early-type spirals. We have performed several tests to 
establish that this result is not due to observational effects: bars are 
visible just as easily in early-type bulges as in bulges of late type 
galaxies. Also, deprojection cannot cause the larger fraction of elongated 
features in late-type spirals.

The result could be explained if bulges of late-type spiral galaxies are 
formed primarily through secular evolution of bars, while this would not 
be the case for earlier-type bulges. Since however it is still unclear what 
the morphology of bulges that have been created through secular evolution, 
it is not possible at this stage to give more detailed conclusions. To better 
understand the formation process of bulges, it is important that we first 
obtain a better understanding of the process of secular evolution.

\section*{Acknowledgement}
We thank Michael Merrifield for very interesting and useful discussions. 
The data presented in this paper were obtained from the Multimission Archive 
at the Space Telescope Science Institute (MAST). STScI is operated by the 
Association of Universities for Research in Astronomy, Inc., under NASA 
contract NAS5-26555. Support for MAST for non-HST data is provided by 
the NASA Office of Space Science via grant NAG5-7584 and by other grants 
and contracts. Much of the analysis was performed using {\small \sc Iraf} 
which is distributed by NOAO.


\setcounter{figure}{6}
\begin{figure*}
\label{fig:profiles}
\resizebox{\hsize}{!}{
\includegraphics{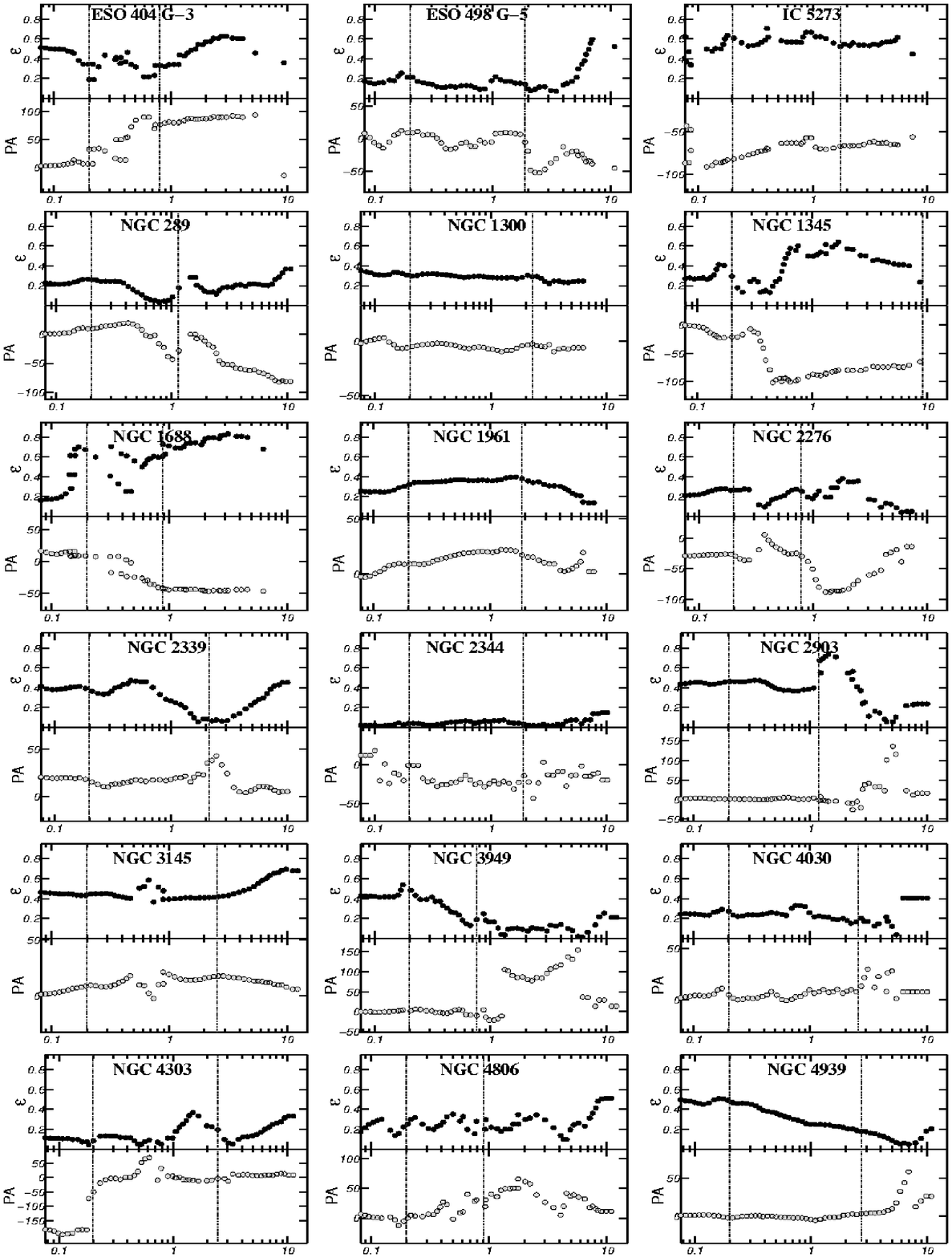}}
\caption{Deprojected ellipticity (top) and PA (bottom) profiles for our
late-type spirals. On the horizontal axis, the radius in arc seconds is
given, and the dashed vertical lines indicate the inner and outer bulge 
radii. The inner bulge radius is defined as the region where ellipse 
fitting can be unreliable (e.g. Peletier et al. 1990; Rest et al. 2001) 
i.e. $\simeq 0.2''$.}
\end{figure*}

\setcounter{figure}{6}
\begin{figure*}
\resizebox{\hsize}{!}{
\includegraphics{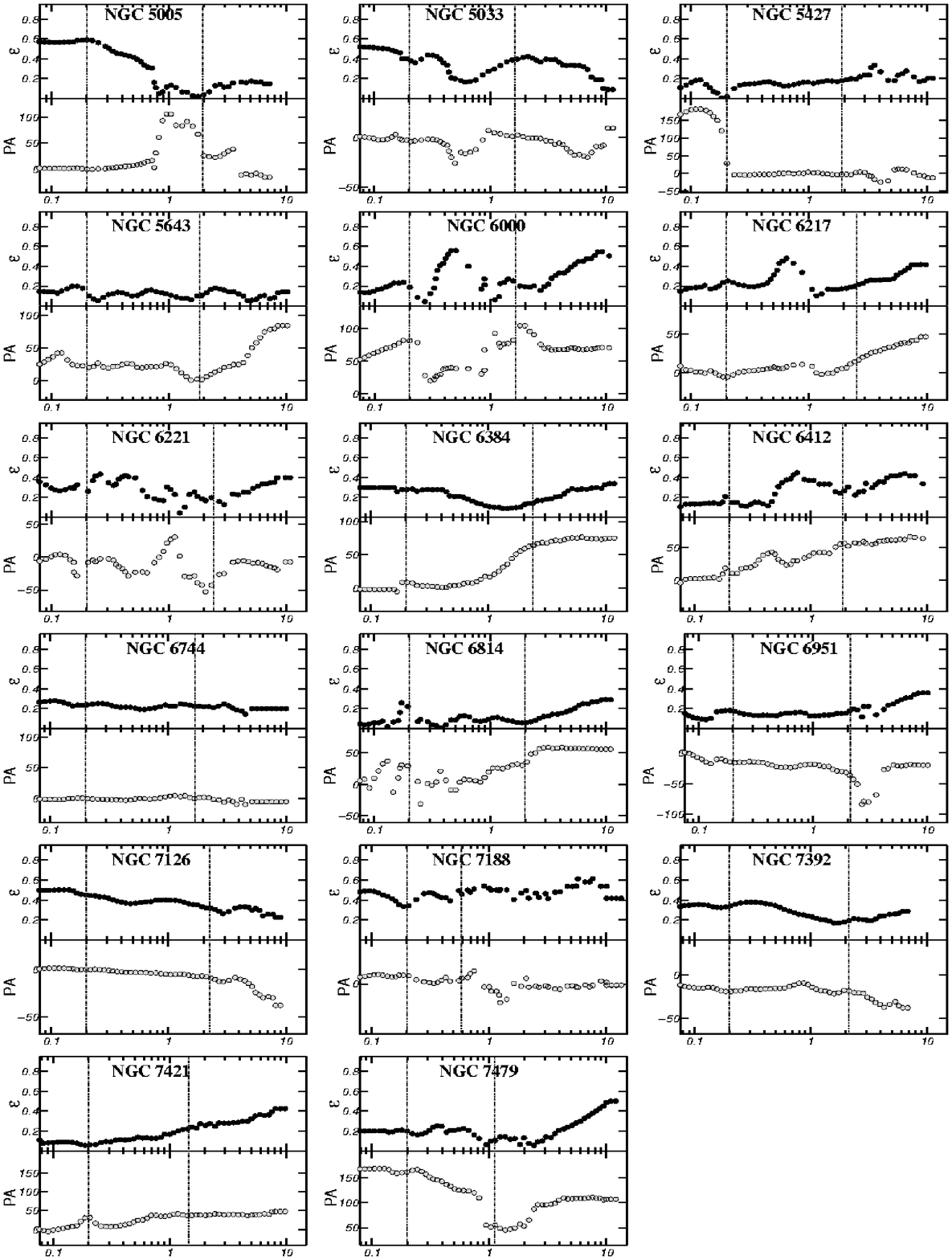}}
\caption{Continued.}
\end{figure*}


\setcounter{figure}{7}
\begin{figure*}
\label{fig:residuals}
\resizebox{\hsize}{!}{
\includegraphics{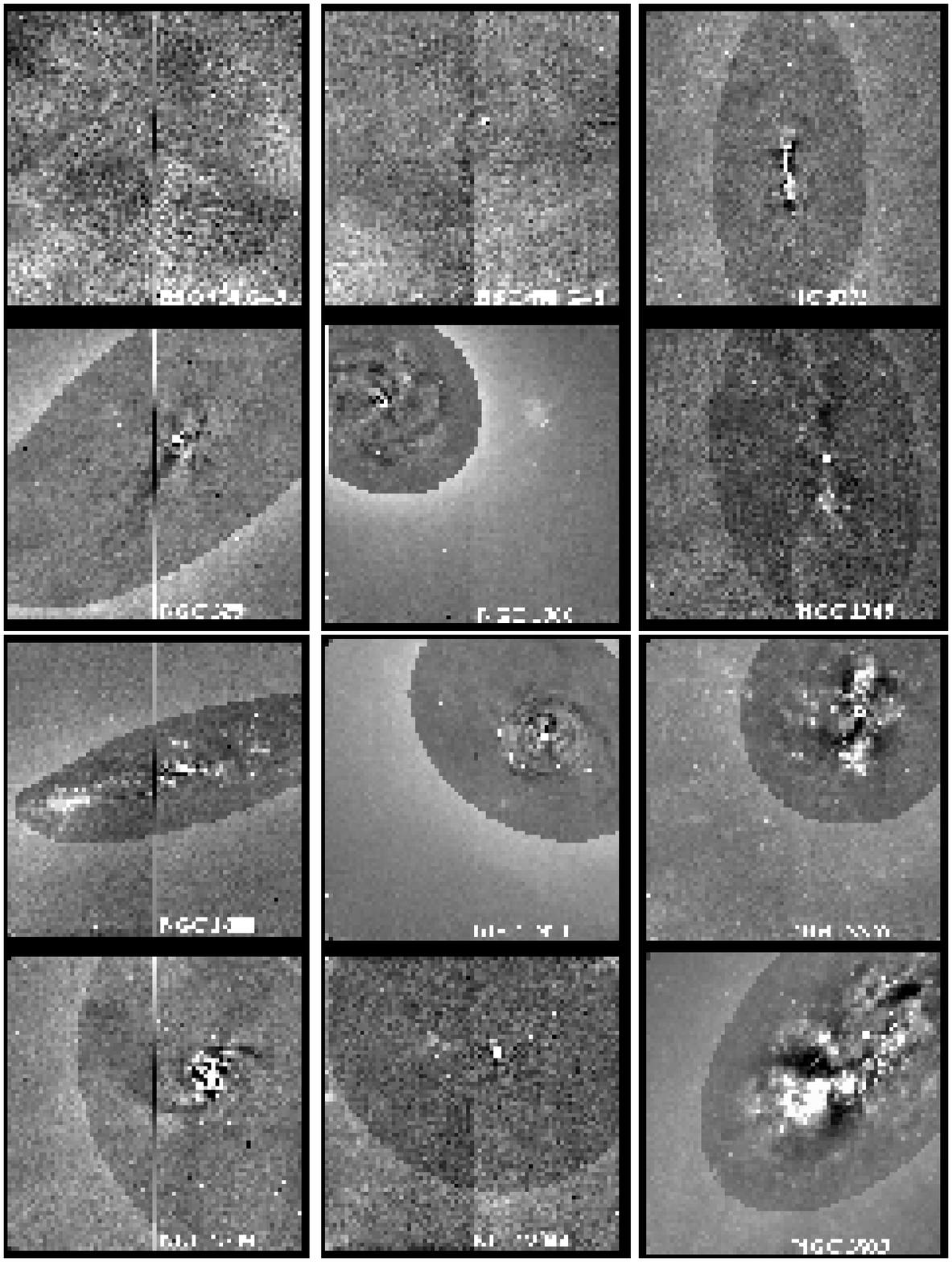}}
\caption{HST/NICMOS (NIC2) H-band residual images 
(original image - model image generated from ellipse fitting). The 
field of view is $19'' \times 19''$ and the pixel size is
$0.075''$. The circular ring marks the estimated centre of 
the galaxy, and its size corresponds to the inner bulge radius.}
\end{figure*}

\setcounter{figure}{7}
\begin{figure*}
\resizebox{\hsize}{!}{
\includegraphics{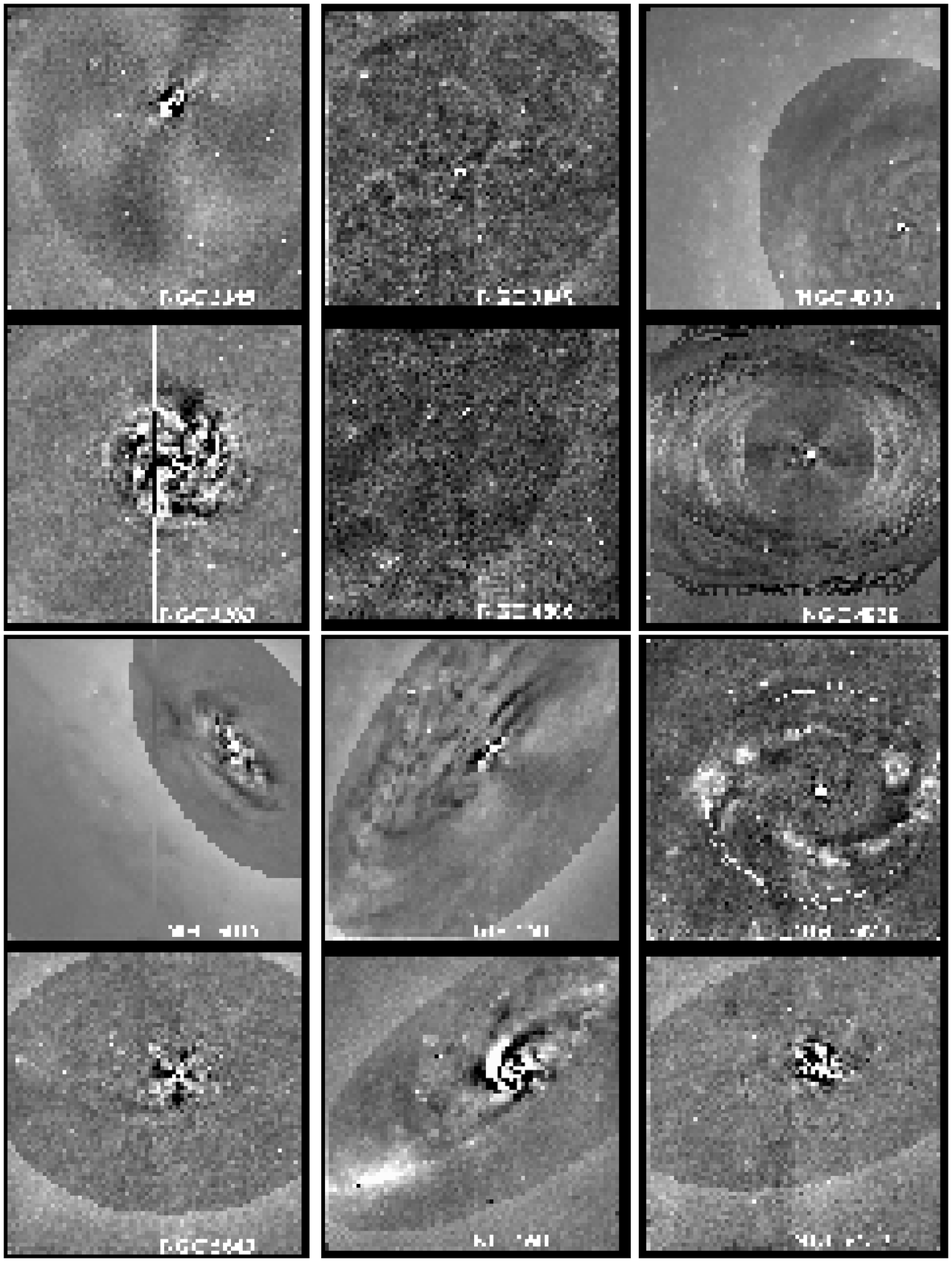}}
\caption{Continued.}
\end{figure*}

\setcounter{figure}{7}
\begin{figure*}
\resizebox{\hsize}{!}{
\includegraphics{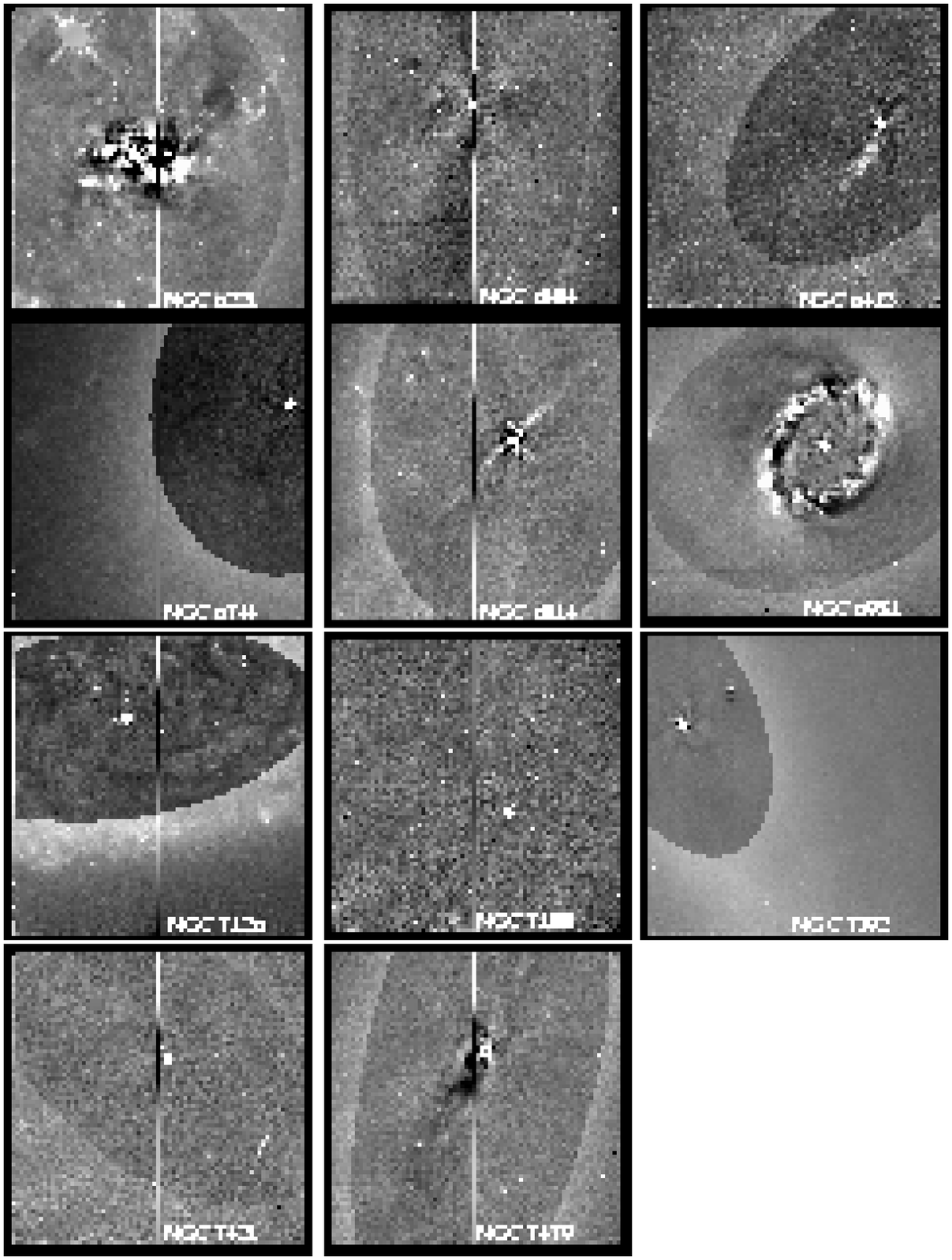}}
\caption{Continued.}
\end{figure*}

\end{document}